\documentclass{llncs}
\usepackage{graphicx}
\usepackage{todonotes}

\begin{document}
\title{Analysis of Schema.org Usage in Tourism Domain}
%
%
\author{Boran Taylan Balc{\i} \and Umutcan \c{S}im\c{s}ek
\and Elias K\"{a}rle \and Dieter Fensel}
%
%
%
\institute{Semantic Technology Institute Innsbruck, Department of Computer Science, University of Innsbruck, Technikerstrasse 21a, 6020 Innsbruck, Austria\\
\email{\{boran.balci, umutcan.simsek, elias.kaerle, dieter.fensel\}@sti2.at}
}

\maketitle              

\begin{abstract}
Schema.org is an initiative founded in 2011 by the four-big search engine Bing, Google, Yahoo!, and Yandex. The goal of the initiative is to publish and maintain the schema.org vocabulary, in order to facilitate the publication of structured data on the web which can enable the implementation of automated agents like intelligent personal assistants and chatbots. In this paper, the usage of schema.org in tourism domain between years 2013 and 2016 is analysed. The analysis shows the adoption of schema.org, which indicates how well the tourism sector is prepared for the web that targets automated agents. The results have shown that the adoption of schema.org type and properties is grown over the years. While the US is dominating the annotation numbers, a drastic drop is observed for the proportion of the US in 2016. Poorly rated businesses are encountered more in 2016 results in comparison to previous years.

\keywords{schema.org, e-tourism, annotation}
\end{abstract}
\section{Introduction}
The schema.org vocabulary has been enabling publication of structured data on the web since 2011. Many businesses have included structured data publication to their online communication strategy, mainly for improving their visibility on search engines that utilize schema.org annotations for bringing better search results. 
In this paper, the analysis is conducted to see the development of the schema.org usage in the tourism domain over the years. The analysis will give us a better understanding of the most demanded types and properties as well as the change of the adoption level within a given time span. The analysis is conducted on the Web Data Commons (WDC) datasets \cite{Meusel2014}, extracted between 2013 and 2016, on 12 diﬀerent tourism related types and their properties. Additionally, a preliminary analysis has been made to see the adoption of the types and properties introduced by the Hotel Extension \cite{karle2017extending}. This extension was published in August 2016; therefore, the analysis is only made for the timespan between August and October 2016.
The relevant data published as N-Quads\footnote{https://www.w3.org/TR/n-quads/} by WDC is loaded to a triple store. Afterwards, a set of SPARQL\footnote{https://www.w3.org/TR/rdf-sparql-query/} queries are run against the triple store for aggregation of the data. Additional post-processing techniques such as Entity Reconciliation\footnote{https://en.wikipedia.org/wiki/Record\_linkage} (ER) and Reverse Geocoding\footnote{https://developers.google.com/maps/documentation/geocoding/start} (RG) applied for the country analysis. The remainder of this paper is structured as follows: In Section \ref{sec:RelatedWork} literature review is explained. Section \ref{sec:Analysis} describes the analysis framework whereas Section \ref{sec:Results} discusses the results. The summary and the future work are explained in Section \ref{sec:Conclusion}.

\section{Related Work}
\label{sec:RelatedWork}
Several analyses of schema.org usage on the web exist in the literature. The study in \cite{10.1007/978-3-319-11955-7_31} describes an implementation of structured data as a CMS extension. The work states that 8.63\% increase in visitors of DMO (Destination Management Organization) Innsbruck website is seen after schema.org implementation in a specific time interval which shows the importance of the structured data in tourism. The analysis made by \cite{10.1007/978-3-319-03973-2_48} indicates that only 5\% of Austrian hotels implement annotations on their web pages whereas our work compares country annotations in a quantitative manner.
The work in \cite{Guha2016} gives some statistics about the most frequently used types between 2011 and 2015. The statistics obtained from 10 billion web pages show that schema.org usage in 2015 increased from 22\% to 31.3\%. In our study, the analysis focuses the usage in a qualitative manner for tourism related types.
\cite{Meusel:2015:WSA:2797115.2797124} makes an analysis which investigates the evolution of the vocabulary alongside the adoption. They show that the evolution works two-ways, (1) users adapt to the changes in the vocabulary well (2) users use non-existent types and properties in their annotations, which inﬂuences the evolution of the vocabulary. Additionally, they point out that around half of the types in schema.org vocabulary is not used at all. The study used the statistics to show the development of schema.org vocabulary while our study provides more specific statistics based on tourism related types in both qualitative and quantitative way.
The closest related work to our work is \cite{Kärle2016}.  Our analysis is an extension to which only covered the Hotel annotations. This paper extends the previous work in the following ways: (a) our study covers several tourism related types and their properties (b) the country based analysis is more accurate with the post-process methods (c) naturally, this analysis covers a longer time span.

\section{Analysis Framework}
\label{sec:Analysis}
\begin{figure}
\centering
\includegraphics[width=0.7\textwidth]{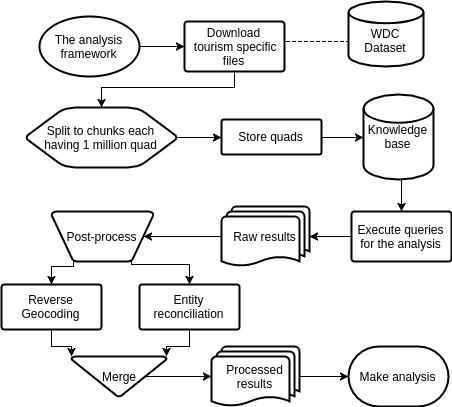}
\caption{The flow diagram of the framework}
\label{fig:framework}
\end{figure}

The data between 2013 and 2016 is analysed based on the classes that are relevant to the tourism domain. These classes can be listed as follows: Airport, Event, Hotel, LakeBodyOfWater, LandmarksOrHistoricalBuildings, LocalBusiness, Mountain, Museum, Park, Restaurant, RiverBodyOfWater and SkiResort. For each year, the total amount of triples vary; in 2013 1.2 billion, in 2014 622 million, in 2015 1.1 billion, in 2016 2.1 billion of triples are processed\footnote{More than 65\% of 2013 annotations come from citysearch.com. This domain was not crawled abundantly in other years which caused the drop on 2014.}. The workflow of the framework is shown in Figure \ref{fig:framework}.

ER is used to reconcile country names in different languages to unified one in order to make an accurate aggregation. OpenRefine\footnote{http://openrefine.org/}’s kNN (k-Nearest Neighbour)\footnote{https://en.wikipedia.org/wiki/K-nearest\_neighbors\_algorithm} method is adopted in the framework with the similarity metric as Levenshtein distance\footnote{https://en.wikipedia.org/wiki/Edit\_distance}. The results are combined with the output of RG which is used to fetch country names.
The analysis is conducted regarding several aspects which are considered essential to understand the schema.org adoption from the tourism point of view. Therefore, a set of queries are prepared and executed, in order to supply results for further deduction. The queries are chosen with respect to following perspectives:
\begin{itemize}
\item Count of all types: All tourism related type instances are fetched along with other types e.g., PostalAddress, ImageObject. 
\item Count of tourism related types’ properties: Tourism related type instances and their properties are fetched to observe the change in frequencies over the years. The wellness of the classes is determined by Mean Squared Error (MSE)\footnote{https://en.wikipedia.org/wiki/Mean\_squared\_error}.
\item Adoption of Hotel extension: Additional instances of the classes from the Hotel Extension\footnote{http://schema.org/docs/hotels.html} are fetched which are introduced in August 2016 as schema.org 3.1.
\item Count of aggregate ratings: The values of aggregateRating\footnote{http://schema.org/aggregateRating} property are fetched and normalized to 1-5 range.
\item Count of Pay-Level Domains(PLDs): Pay-Level Domains(PLDs)\footnote{http://webdatacommons.org/structureddata/vocabulary-usage-analysis/} are obtained by querying the context of quads.
\item Count of countries: addressCountry\footnote{http://schema.org/addressCountry} values are processed with ER and RG.
\end{itemize}	

In the following section \ref{sec:Results}, the results will be discussed.

\section{Results}
\label{sec:Results}
The observation shows that the data distribution among 12 tourism related types is disproportional. The annotations of natural beauties are found very rare in comparison to business related schema.org types such as Restaurant, Hotel, and LocalBusiness. In order to assess how well the annotation of these types are implemented, MSE values are used. The result shows that the quality of annotations generally increased over the years since MSE values decrease or remain same.

It is observed that majority of the entities are located in the US. However, over the years, the distribution of annotations is slightly shifting to other countries. For instance, in Hotel annotations, for 2013 and 2014 the ratio of the US are 79\%, for 2015 72\% and for 2016 the number decreases drastically to 28.4\% which evidently shows that the schema.org adoption is increased by other countries over time. 

\begin{figure}
\centering
\includegraphics[width=0.7\textwidth]{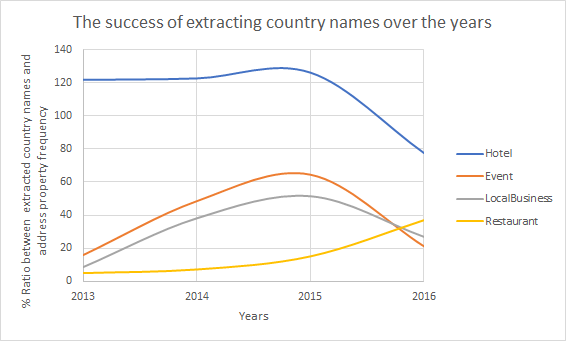}
\caption{Reconciliation ratio over address property frequency}
\label{fig:successratio2}
\end{figure}

One other important aspect is how well the implementation of geo and address property is made over the years. Figure \ref{fig:successratio2} shows that Hotel annotations contain address or geolocation information to extract the country names whereas the success ratio of other classes could not exceed 70\%. As a result, the implementation of geo and address properties remain sparse in other classes.

Rating values showed an increase for the low values from 2015 to 2016. The study in \cite{park2015asymmetric} analyses 5090 Restaurant reviews to find out the potential effect of ratings. The result shows that consumers appreciate extreme ratings especially bad ratings more than medium ratings. Our findings support this result.
The analysis of the Hotel Extension is made for the 3 months period between August and October 2016. The newly introduced types Campground and HotelRoom have 716 and 117 annotations respectively. Type Room has been used 3339 times. The adopters seem to use LocationFeatureSpeciﬁcation for amenities almost 7000 times. Signiﬁcant usage of the hasAmenity property is observed for Room type, 17000 times to be precise. A quick search we made has shown that hasAmenity property was proposed by an early extension attempt\footnote{https://www.w3.org/wiki/WebSchemas/LodgingExtensions} in 2013. The analysis shows that, within the 3 months period, some classes introduced by the extension started to be used. However, most of the classes have no occurrence in the WDC datasets.

\section{Conclusion}
\label{sec:Conclusion}
In this paper, an analysis conducted with a tourism related subset of schema.org from diﬀerent perspectives (i.e., types, properties, geo data, PLD, and aggregated rating) has been presented. It is observed that the adoption has become better over the years. However, the values of geo and addressCountry properties were insuﬃcient to determine country names for the instances, except for the Hotel type. The newly introduced classes in the hotel extension are used in a small amount of annotations. A signiﬁcant increase of low ratings is seen from 2015 to 2016. Even though, the quality of schema.org annotations is improved, there are still issues with important properties such as address. Nevertheless, touristic service providers not only in the US, but also in other countries is increasingly getting ready for the automated agents on the web.
As a future work, PLDs can be analysed in more detail to observe the improvement in annotation pattern. The analysis of 2017 WDC data dump would show better results for the hotel extension since 2016 data dump is extracted in an early stage of the extension. The extended results are under the link http://btbalci.sti2.at/enter2018/.

%
%

\bibliographystyle{apalike}
\bibliography{references}

\end{document}